\documentclass[letterpaper]{elsart5p}
 \usepackage{graphicx}
\usepackage{amssymb}
\begin{document}

\begin{frontmatter}

\title{Polarized neutron scattering studies of the kagom\'e lattice antiferromagnet KFe$_{3}$(OH)$_{6}$(SO$_{4}$)$_{2}$}

\author[ad1]{K.~Matan\thanksref{pread}\corauthref{cor1}},
\ead{kmatan@issp.u-tokyo.ac.jp}
\author[ad1]{J.~S.~Helton},
\author[ad2]{D.~Grohol},
\author[ad3]{D.~G.~Nocera},
\author[ad4]{S.~Wakimoto},
\author[ad4]{K.~Kakurai},
\author[ad1]{Y.~S.~Lee}
\address[ad1]{Department of Physics, Massachusetts Institute of Technology, Cambridge, MA 02139, USA}
\address[ad2]{The Dow Chemical Company, Core R\&D, Midland, MI 48674, USA}
\address[ad3]{Department of Chemistry, Massachusetts Institute of Technology, Cambridge, MA 02139, USA}
\address[ad4]{Quantum Beam Science Directorate, Japanese Atomic Energy Agency, 2-4 Shirakata Shirane, Tokai, Naka, Ibaraki 319-1195, Japan}
\thanks[pread]{Present address: Neutron Science Laboratory, Institute for Solid State Physics, University of Tokyo, 5-1-5 Kashiwanoha, Kashiwa, Chiba 277-8581, Japan.}
\corauth[cor1]{Corresponding author.}

\begin{abstract}
We report polarized neutron scattering studies of spin-wave excitations and spin fluctuations in the S=5/2 kagom\'e lattice antiferromagnet KFe$_{3}$(OH)$_{6}$(SO$_{4}$)$_{2}$ (jarosite). Inelastic polarized neutron scattering measurements at 10 K on a single crystal sample reveal two spin gaps, associated with in-plane and out-of-plane excitations. The polarization analysis of quasi-elastic scattering at 67 K shows in-plane spin fluctuations with XY symmetry, consistent with the disappearance of the in-plane gap above the N\'eel temperature T$_N$ = 65 K. Our results suggest that jarosite is a promising candidate for studying the 2D XY universality class in magnetic systems.
\end{abstract}

\end{frontmatter}

\section{Introduction}

\begin{figure}
\center
\includegraphics[width=3in]{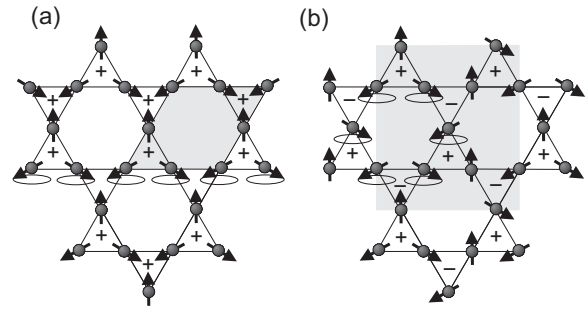}
\caption{Examples of ground state spin configurations of the kagom\'e lattice antiferromagnet.  The magnetic unit cell is shown by the shaded areas.  The positive and negative signs indicated spin chirality\cite{grohol_chiral}.  The dotted loops illustrate the zero energy excitations as described in the text for (a) q=0 and (b) $\sqrt{3}\times\sqrt{3}$ states.}\label{fig1}
\end{figure}

Collective behavior in strongly correlated systems often leads to unconventional ground states.  One example of such systems is geometrically frustrated spins, where the incompatibility between the topology of an underlying lattice and spin interactions prevents the system from selecting a unique ground state, prompting the presence of novel spin dynamics.  The kagom\'e lattice antiferromagnet is a highly frustrated two-dimensional lattice, being comprised of corner-sharing triangles.  One of the hallmarks of this system is the ``zero energy modes'', which result from the highly degenerate, but connected, ground state manifold\cite{harris}.  The only constraint for the ground state is that the spins on each triangle be oriented 120$^\circ$ relative to each other.  The loops at the tips of the spins in Fig.~\ref{fig1} illustrate rotations of two of the spin sublattices about the axis defined by the third spin sublattice, representing the out-of-plane excitation with no energy cost (the 120$^\circ$ arrangement on each triangle is maintained). The spins on different parallel chains in the $q=0$ structure (Fig.~\ref{fig1}(a)) and different hexagons in the $\sqrt{3}\times\sqrt{3}$ structure (Fig.~\ref{fig1}(b)) can be excited independently, resulting in the non-dispersive excitation. 

%===================================================
%=====================================================

One ideal realization of the kagom\'e lattice antiferromagnet is jarosite KFe$_{3}$(OH)$_{6}$(SO$_{4}$)$_{2}$.  Fe$^{3+}$ magnetic ions with $S=5/2$ located inside tilted octahedral cages formed by six oxygen atoms reside at each corner of the corner-sharing triangles that form the perfect kagom\'e planes. Due to the presence of the Dzyaloshinskii-Moriya (DM) interaction, the spins order in a $q=0$ structure (Fig.~\ref{fig1}(a)) with small canting forming an ``umbrella'' spin configuration below the N\'eel temperature T$_N=65$ K\cite{grohol_chiral,inami,wills:064430}. The DM interaction is allowed in jarosite because there is no inversion center between the magnetic ions\cite{dzyaloshinskii,moriya}. The spin Hamiltonian is given by:
\begin{equation}
{\cal H}=\sum_{nn} \left[ J_1 \, \textbf{S}_i \cdot \textbf{S}_j \, + \,
\textbf{D}_{ij}\cdot \textbf{S}_i \times \textbf{S}_j \right]+\sum_{nnn}J_2 \, \textbf{S}_i \cdot \textbf{S}_j, \label{eq1}
\end{equation}
where $J_1$ and $J_2$ are the nearest-neighbor and next-nearest-neighbor interactions, respectively, and $\textbf{D}_{ij}=(0, D_p, D_z)$ is the DM vector. The DM interaction lifts the ``zero energy mode'' to a finite energy, making it measurable by neutron scattering. We have previously reported on the first observation of this lifted zero energy mode in jarosite, and determined all relevant spin Hamiltonian parameters\cite{matan}. In this proceeding, we report polarized neutron scattering studies of the lifted zero energy mode, and other spin-wave modes in jarosite.  The polarization of spin fluctuations above $T_N$ is also investigated.

\section{Experimental}

Inelastic polarized neutron scattering measurements were performed on TAS-1 spectrometer at Japan Atomic Energy Agency, Tokai, Japan.  This state-of-the-art instrument utilizes double-focusing Heussler crystals to monochromate and analyze the incident and scattered neutron beams\cite{takeda}.  The beam polarization of more than 90\% is achieved.  The final neutron energy was fixed at 14.7 meV.  Horizontal collimations of open$-80'-$sample$-$open$-$open were employed.  However, due to the small sample, effective horizontal collimations of $48'-18'-\mbox{sample}-25'-240'$ were used to calculate the resolution function.  Pyrolytic graphite filters were placed in the scattered beam to reduce higher-order contaminations. The energy resolution of about 1.2 meV and background of about one count per minute were achieved. A single crystal of jarosite of mass 48 mg was oriented in the (HK0) zone. The sample was cooled by a $^4$He close-cycle displex.

The spin-wave calculations of the spin Hamiltonian (Eq.~\ref{eq1}) show three branches of low energy spin excitations, which correspond to two out-of-plane excitations and one in-plane excitation\cite{yildirim,nishiyama}.  By choosing proper orientations of a polarization vector $\textbf{P}$, which is defined by the direction of a guide field, and wave vector $\textbf{Q}$, one can distinguish between the in-plane and out-of-plane spin excitations.  If $\textbf{P}$ and $\textbf{Q}$ are parallel, both in-plane and out-of-plane magnetic excitations will give rise to scattering intensity in the spin-flip (SF) channel.  On the other hand, if $\textbf{P}$ is perpendicular to $\textbf{Q}$, then the magnetic scattering in the SF channel is due to the excitations that are perpendicular to both $\textbf{Q}$ and $\textbf{P}$, and the magnetic scattering in the non-spin-flip (NSF) channel is due to the excitations parallel to $\textbf{P}$.

%===================================================
\begin{figure}
\center
\includegraphics[width=3in]{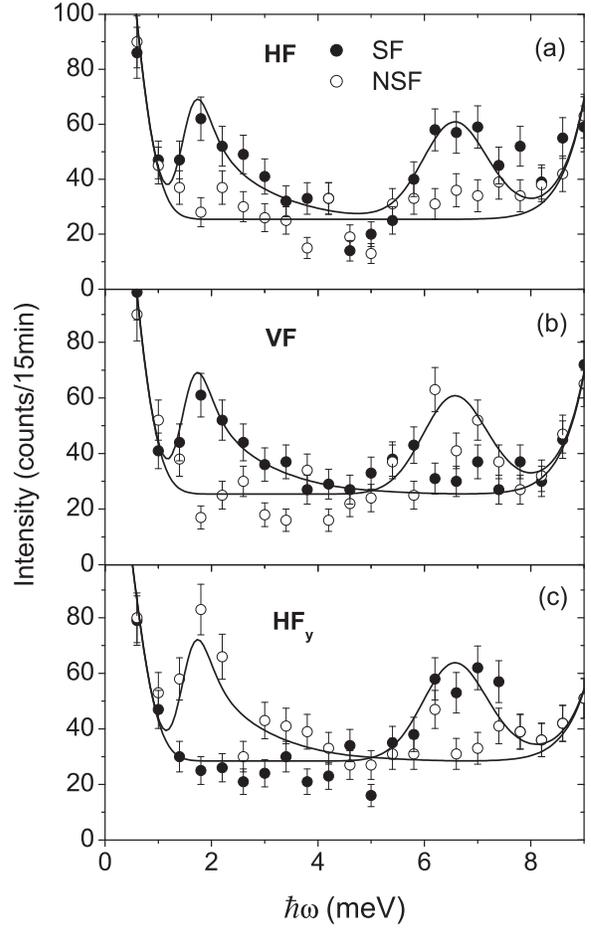}
\caption{Inelastic polarized neutron scattering measurements of spin-wave excitations at the zone center, $\textbf{Q}=(1 0 0)$.  The guide field is (a) parallel to $\textbf{Q}$, (b) perpendicular to $\textbf{Q}$ and scattering plane, and (c) perpendicular to $\textbf{Q}$ but lies within the scattering plane.}\label{fig2}
\end{figure}
%=====================================================

Fig.~\ref{fig2} shows constant-$\textbf{Q}$ scans at (1,0,0), the center of the Brillouin zone ($\Gamma-$point), measured at 10 K.  The observed peaks were fit with narrow Gaussian convoluted with the experimental resolution function assuming the empirical dispersion\cite{matan}.  All fitting parameters except peak intensities were the same as those obtained from the unpolarized neutron scattering measurement reported in Ref.~\cite{matan}. The rise at high energy was due to the contaminations from the main beam at $2\theta=0$. For the guide field along $\textbf{Q}$ (HF) (Fig.~\ref{fig2}(a)), both spin-wave excitations at 2 meV and 7 meV were observed in the SF channel, while there is no magnetic scattering in the NSF channel.  For the guide field perpendicular to $\textbf{Q}$, and perpendicular to the scattering plane (VF) (Fig.~\ref{fig2}(b)), the peak at 2 meV was observed in the SF channel, and the peak at 7 meV was observed in the NSF channel.  On the other hand, if the guide field was perpendicular to $\textbf{Q}$, but lay within the scattering plane (HF$_\textrm{y}$) (Fig.~\ref{fig2}(c)), then the peak at 2 meV was observed in the NSF channel and  the peak at 7 meV was observed in the SF channel.  The equivalent intensities in the VF SF and HF SF channels and the absence of intensity in the VF NSF channel at 2 meV indicate that the 2 meV excitation is predominantly the in-plane excitation.  Similarly, the equivalent intensities in the VF NSF and HF SF channels and the absence of intensity in the VF SF channel at 7 meV indicate that the 7 meV excitation is predominantly the out-of-plane excitation.  These results conclusively show that the 2 meV gap is an in-plane excitation, and the 7 meV gap is an out-of-plane excitation, consistent with the spin-wave analysis.  At the zone center, given the resolution of the instrument, the two out-of-plane modes appear degenerate at around 7 meV, with one of these modes being the lifted zero energy mode.

%===========================================================
\begin{figure}
\center
\includegraphics[width=3in]{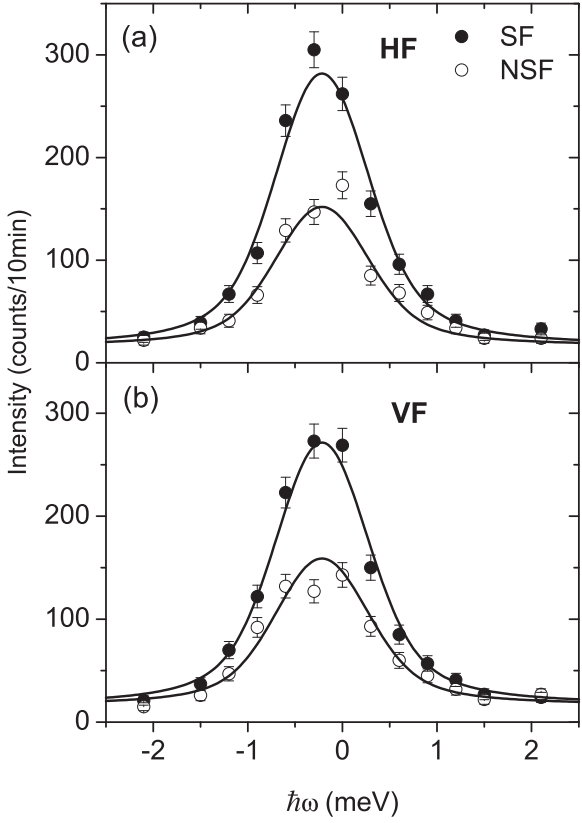}
\vspace{0in} \caption{Quasi-elastic scattering centered at (1 0 0) measured by polarized neutrons at 67 K, when the guide field is parallel to $\textbf{Q}$ (a), and perpendicular to $\textbf{Q}$ and the scattering plane (b).}\label{fig3}
\end{figure}
%===========================================================

The polarization of the spin fluctuations were also studied using polarized neutron scattering.  Fig.~\ref{fig3} shows constant-$\textbf{Q}$ scan at (1 0 0) through a quasi-elastic peak measured at $T=67$ K.  The observed quasi-elastic peak was fit to Lorentzian convoluted with the experimental resolution function.  Backgrounds and peak widths were kept constant for all data sets.  The intensities in the HF SF and VF SF channels indicate that the spin fluctuations are predominantly the in-plane fluctuations, providing the direct evidence for the presence of $XY$ symmetry previously reported in Ref.~\cite{grohol_chiral}.

\section{Summary and discussion}

In jarosite, the spin gaps are a result of the antisymmetric DM interaction.  To the leading order, the analytic expressions show that the in-plane gap is proportional to $|D_p|$ while the two out-of-plane gaps are proportional to $\sqrt{J_1D_z}$\cite{matan,yildirim}.  Therefore, despite the similar $D_p$ and $D_z$, the out-of-plane gap is significantly larger than the in-plane gap.  Effectively, $D_p$ and $D_z$ give rise to an easy-axis and easy-plane anisotropies, respectively. $D_z$, which confines the spins in the plane, breaks the chiral symmetry, giving rise to the $q=0$ arrangement with positive chirality, while $D_y$ breaks the rotational symmetry around the $c-$axis, resulting in the Ising-type of ordering below $T_N$\cite{elhajal}.  However, above $T_N$ only the in-plane rotational ($XY$) symmetry is restored, while the chiral symmetry remains broken up to higher temperature\cite{grohol_chiral}.  This result immediately suggests the existence of the out-of-plane gap above $T_N$.  In fact, our polarized neutron scattering around the quasi-elastic peak above $T_N$ exhibits the in-plane-only spin fluctuations, which imply that the out-of-plane gap remains at a finite energy.  Therefore, in the critical regime, KFe$_3$(OH)$_6$(SO$_4$)$_2$, whose interlayer coupling is negligibly small\cite{grohol_chiral}, is a promising candidate for studying the 2D $XY$ universality class in magnetic systems.
\\
\\
\textbf{Acknowledgement}
\\
\\
We thank Japanese Atomic Energy Agency for the provision of neutron beam time.  This work was supported by the NSF under Grant no. DMR 0239377, and in part by the MRSEC program under Grant no. DMR 02-13282.

\bibliographystyle{elsart-num}
\bibliography{KFe3J_polarized_neutrons}

\end{document}